\documentclass[hidelinks,11pt]{article}

\usepackage{amsfonts,amsthm,xspace,graphicx,relsize,bm,mathtools,xcolor,amsmath}
\usepackage{mathrsfs}
\usepackage[pagebackref]{hyperref}

\renewcommand{\backref}[1]{}

\renewcommand{\backrefalt}[4]{%
\ifcase #1
\or $^{#2}$
\else $^{#2}$
\fi}
\usepackage{kpfonts}
\usepackage[margin=1in]{geometry}
\hypersetup{
    colorlinks,
    linkcolor={blue!100!black},
    citecolor={blue!100!black},
}

\newtheorem{theorem}{Theorem}[section] 

\newtheorem{lemma}[theorem]{Lemma}

\newtheorem{corollary}[theorem]{Corollary}
\newtheorem{definition}[theorem]{Definition}

\newcommand{\C}{\ensuremath{\mathbb{C}}}
\newcommand{\Se}{\ensuremath{\mathcal{S}}}

\newcommand{\De}{\ensuremath{\mathcal{D}}}

\newcommand{\A}{\ensuremath{\mathcal{A}}}

\newcommand{\N}{\ensuremath{\mathbb{N}}}

\newcommand{\F}{\ensuremath{\mathbb{F}}}

\newcommand{\R}{\ensuremath{\mathbb{R}}}

\newcommand{\E}{\mathcal{E}}
\newcommand{\M}{\mathcal{M}}

\newcommand{\Cc}{{\mathcal C}} 
\newcommand{\eps}{\varepsilon}
\newcommand{\ket}[1]{|#1\rangle}
\newcommand{\bra}[1]{\langle#1|}

\newcommand{\ip}[2]{\langle #1 | #2 \rangle}
\newcommand{\ketbra}[2]{|#1\rangle\! \langle #2|}
\newcommand{\braketbra}[3]{\langle #1|#2| #3 \rangle}
\newcommand{\Tr}{\mbox{\rm Tr}}
\newcommand{\poly}{\mbox{\rm poly}}
\newcommand{\PEX}{\mbox{\rm PEX}}
\newcommand{\MAJ}{\mbox{\rm MAJ}}
\newcommand{\QPEX}{\mbox{\rm QPEX}}

\newcommand{\MQ}{\mbox{\rm MQ}}
\newcommand{\QMQ}{\mbox{\rm QMQ}}
\newcommand{\QAEX}{\mbox{\rm QAEX}}
\newcommand{\AEX}{\mbox{\rm AEX}}
\newcommand{\err}{\mbox{\rm err}}
\newcommand{\opt}{\mbox{\rm opt}}
\newcommand{\Id}{\ensuremath{\mathop{\rm Id}\nolimits}}

\newcommand{\Exp}{\mathbb{E}}
\def\01{\{0,1\}}
\newcommand{\tvect}[2]{%
  \ensuremath{\Bigl(\negthinspace\begin{smallmatrix}#1\\#2\end{smallmatrix}\Bigr)}}

\newenvironment{proofsketch}
{\noindent \emph{Proof sketch.}}
{{\hfill $\Box$}\\
 \smallskip}

\title{A Survey of Quantum Learning Theory}
\author{Srinivasan Arunachalam\thanks{QuSoft, CWI, Amsterdam, the Netherlands.
Supported by ERC Consolidator Grant QPROGRESS 615307.}
\and
Ronald de Wolf\thanks{QuSoft, CWI and University of Amsterdam, the Netherlands.
Partially supported by ERC Consolidator Grant QPROGRESS 615307.}
}
\date{
}
\begin{document}
\maketitle

\begin{abstract}
This paper surveys quantum learning theory: the theoretical aspects of machine learning using quantum computers. We describe the main results known for three models of learning: exact learning from membership queries, and Probably Approximately Correct (PAC) and agnostic learning from classical or quantum examples.
\end{abstract}

\section{Introduction}

Machine learning entered theoretical computer science in the 1980s with the work of Leslie Valiant~\cite{valiant:paclearning}, who introduced the model of ``Probably Approximately Correct'' (PAC) learning, building on earlier work of Vapnik and others in statistics, but adding computational complexity aspects. 
This provided a mathematically rigorous definition of what it means to (efficiently) learn a target concept from given examples.
In the three decades since, much work has been done in computational learning theory: some efficient learning results, many hardness results, and many more models of learning. We refer to~\cite{kearnss&vazirani:learningbook,anthony&bartlett:learningbook,shwartz&david:learningbook} for general introductions to this area.
In recent years practical machine learning has gained an enormous boost from the success of deep learning in important big-data tasks like image recognition, natural language processing, and many other areas; this is theoretically still not very well understood, but it often works amazingly~well.

Quantum computing started in the 1980s as well, with suggestions for analog quantum computers by Manin~\cite{manin:radio,manin:qc},
Feynman~\cite{feynman:simulating,feynman:qmc}, and Benioff~\cite{benioff:hamiltonian}, and reached more digital ground with Deutsch's
definition of a universal quantum Turing machine~\cite{deutsch:uqc}.
The field gained momentum with Shor's efficient quantum algorithms~\cite{shor:factoring} for factoring integers and computing discrete logarithms (which between them break much of today's public-key cryptography), and has since blossomed into a major area at the crossroads of physics, mathematics, and computer~science.

Given the successes of both machine learning and quantum computing, combining these two strands of research is an obvious direction.
Indeed, soon after Shor's algorithm, Bshouty and Jackson~\cite{bshouty:quantumpac} introduced a version of learning from \emph{quantum} examples, which are quantum superpositions rather than random samples. They showed that Disjunctive Normal Form (DNF) can be learned efficiently from quantum examples under the uniform distribution; efficiently learning DNF from uniform \emph{classical} examples (without membership queries) was and is an important open problem in classical learning theory. Servedio and others~\cite{atici&servedio:qlearning,atici&servedio:testing,servedio&gortler:equivalencequantumclassical} studied upper and lower bounds on the number of quantum membership queries or quantum examples needed for learning, and more recently the authors of the present survey obtained optimal bounds on quantum sample complexity~\cite{arunachalam:optimalpaclearning}.

Focusing on specific learning problems where quantum algorithms may help, A{\"{\i}}meur et al.~\cite{aimeur:mlinquantumworld,aimeur:qspeedup} showed quantum speed-up in learning contexts such as clustering via minimum spanning tree, divisive clustering, and $k$-medians, using variants of Grover's search algorithm~\cite{grover:search}. In the last few years there has been a flurry of interesting results applying various quantum algorithms (Grover's algorithm, but also phase estimation, amplitude amplification~\cite{bhmt:countingj}, and the HHL algorithm for solving well-behaved systems of linear equations~\cite{hhl:lineq}) to specific machine learning problems. Examples include
Principal Component Analysis~\cite{lmp:pca},
support vector machines~\cite{rms:svm},
$k$-means clustering~\cite{lmp:supunsup}, quantum recommendation systems~\cite{kerenidis&prakash:quantumrecommendation},
and work related to neural networks~\cite{wiebe:quantumperceptronmodels,wiebe:quantumdeeplearning}.
Some of this work---like most of application-oriented  machine learning in general---is heuristic in nature rather than mathematically rigorous.
Some of these new approaches are suggestive of exponential speed-ups over classical machine learning, though one has to be careful about the underlying assumptions needed to make efficient quantum machine learning possible: in some cases these also make efficient \emph{classical} machine learning possible. Aaronson~\cite{aaronson:fineprint} gives a brief but clear description of the issues.
These developments have been well-served by a number of recent survey papers~\cite{schuldea:introqml,adcockea:qml,biamonteea:qml} and even a book~\cite{wittek:qml}.

In contrast, in this survey
we focus on the theoretical side of quantum machine learning: quantum learning theory.\footnote{The only other paper we are aware of to survey quantum learning theory is an unpublished manuscript by Robin Kothari from 2012~\cite{kothari:qclearn} which is much shorter but partially overlaps with ours; we only saw this after finishing a first version of our survey.}
We will describe (and sketch proofs of) the main results that have been obtained in three main learning models. These will be described in much more detail in the next sections, but below we give a brief preview.

\paragraph{Exact learning.}
In this setting the goal is to learn a target concept from the ability to interact with it.  For concreteness, we focus on learning
target concepts that are Boolean functions: the target is some unknown $c:\01^n\to\01$ coming from a known concept class $\Cc$ of functions,\footnote{Considering concept classes over $\01^n$ has the advantages that the $n$-bit $x$ in a labeled example $(x,c(x))$ may be viewed as a ``feature vector''. This fits naturally when one is learning a type of objects characterized by patterns involving $n$ features that each can be present or absent in an object, or when learning a class of $n$-bit Boolean functions such as small decision trees, circuits, or DNFs. However, we can (and sometimes do) also consider concepts $c:[N]\to~\01$.} and our goal is to identify $c$ exactly, with high probability, using \emph{membership queries} (which allow the learner to learn $c(x)$ for $x$ of his choice).
If the measure of complexity is just the number of queries, the main results are that quantum exact learners can be polynomially more efficient than classical, but not more. If the measure of complexity is \emph{time}, then under reasonable complexity-theoretic assumptions some concept classes can be learned much faster from quantum membership queries (i.e., where the learner can query $c$ on a superposition  of $x$'s) than is possible classically.

\paragraph{PAC learning.}
In this setting one also wants to learn an unknown $c:\01^n\to\01$ from a known concept class $\Cc$, but in a more passive way than with membership queries: the learner receives several \emph{labeled examples} $(x,c(x))$, where $x$ is distributed according to some unknown probability distribution~$D$ over~$\01^n$. The learner gets multiple i.i.d.\ labeled examples. From this limited ``view'' on $c$, the learner wants to generalize, producing a \emph{hypothesis} $h$ that probably agrees with~$c$ on ``most'' $x$, \emph{measured according to the same~$D$}. This is the classical Probably Approximately Correct (PAC) model. In the quantum PAC model~\cite{bshouty:quantumpac}, an example is not a random sample but a \emph{superposition} $\sum_{x\in\01^n}\sqrt{D(x)}\ket{x,c(x)}$.
Such quantum examples can be useful for some learning tasks with a fixed distribution~$D$ (e.g., uniform $D$) but it turns out that in the usual distribution-independent PAC model, quantum and classical sample complexity are equal up to constant factors, for every concept class $\Cc$.
When the measure of complexity is \emph{time}, under reasonable complexity-theoretic assumptions, some concept classes can be PAC learned much faster by quantum learners (even from classical examples) than is possible classically.

\paragraph{Agnostic learning.}
In this setting one wants to approximate a distribution on $\01^{n+1}$ by finding a good hypothesis $h$ to predict the last bit from the first $n$ bits. A ``good'' hypothesis is one that is not much worse than the best predictor available in a given class $\Cc$ of available hypotheses. The agnostic model has more freedom than the PAC model and allows to model more realistic situations, for example when the data is noisy or when no ``perfect'' target concept exists. Like in the PAC model, it turns out quantum sample complexity is not significantly smaller than classical sample complexity in the agnostic model.

\bigskip

\paragraph{Organization.}
The survey is organized as follows.
In Sections~\ref{sec:introtoquantum} and~\ref{sec:introlearningmodels} we first introduce the basic notions of quantum and learning theory, respectively.  In Section~\ref{sec:querycomplexityandsamplecomplexity} we describe the main results obtained for information-theoretic measures of learning complexity, namely query complexity of exact learning and sample complexities of PAC and agnostic learning.  In Section~\ref{sec:timecomplexity} we survey the main results known about \emph{time} complexity of quantum learners. We conclude in Section~\ref{sec:summaryandoutlook} with a summary of the results and some open questions for further research.

\section{Introduction to quantum information}
\label{sec:introtoquantum}
\subsection{Notation}
For a general introduction to quantum information and computation we refer to~\cite{nielsen&chuang:qc}. In this survey, we assume familiarity with the following notation. Let $\ket{0}=\tvect{1}{0}$ and $\ket{1}=\tvect{0}{1}$ be the standard basis states for~$\C^2$, the space in which one qubit ``lives''.  
Multi-qubit basis states are obtained by taking tensor products of one-qubit basis states; for example, $\ket{0}\otimes\ket{1}\in\C^4$ denotes a basis state of a 2-qubit system where the first qubit is in state $\ket{0}$ and the second qubit is in state~$\ket{1}$. 
For $b\in \01^k$, we often shorthand $\ket{b_1}\otimes \cdots \otimes \ket{b_k}$  as $\ket{b_1\cdots b_k}$. A $k$-qubit \emph{pure} state $\ket{\phi}$ can be written as $\ket{\phi}=\sum_{i\in \01^k} \alpha_i \ket{i}$ where the $\alpha_i$'s are complex numbers (called \emph{amplitudes}) that have to satisfy $\sum_{i\in \01^k}|\alpha_i|^2=1$. We view $\ket{\phi}$ as a $2^k$-dimensional column vector. The row vector that is its complex conjugate is denoted by~$\bra{\phi}$. An $r$-dimensional \emph{quantum state}~$\rho$ (also called a \emph{density matrix}) is an~$r\times r$ positive semi-definite~(psd) matrix~$\rho$ with trace~1; this can be written (often non-uniquely) as $\rho=\sum_i p_i\ketbra{\phi_i}{\phi_i}$ and hence can be viewed as a probability distribution over pure states~$\ket{\phi_i}$. 

Non-measuring quantum operations correspond to \emph{unitary} matrices $U$, which act by left-multiplication on pure states $\ket{\psi}$ (yielding $U\ket{\psi}$), and by conjugation on mixed states $\rho$ (yielding $U\rho U^{-1}$). 
For example, the $1$-qubit Hadamard transform $H=\frac{1}{\sqrt{2}}\left(
\begin{array}{rr}
1 & 1\\
1 & -1
\end{array}
\right)$ 
corresponds to the unitary map $H:\ket{a}\rightarrow (\ket{0}+(-1)^a\ket{1})/\sqrt{2}$, for $a\in \01$.

To obtain classical information from a quantum state $\rho$, one can apply a \emph{quantum measurement} to~$\rho$. An $m$-outcome quantum measurement, also called a \emph{POVM} (positive-operator-valued measure), is described by a set of positive semi-definite matrices $\{M_i\}_{i\in [m]}$ that satisfy $\sum_{i} M_i= \Id$. When measuring~$\rho$ using this POVM, the probability of outcome~$j$ is given by $\Tr(M_j\rho)$.
 
 \subsection{Query model}
 \label{section:quantumqueryalgorithms}
In the query model of computation, the goal is to compute a Boolean function $f:\01^N\rightarrow \01$ on some input $x\in \01^N$. We are not given $x$ explicitly, instead we are allowed to \emph{query} an oracle that encodes the bits of $x$, i.e., given $i\in [N]$, the oracle returns $x_i$. The cost of a query algorithm is the number of queries the algorithm makes to the oracle. We will often assume for simplicity that $N$ is a power of 2, $N = 2^n$, so we can identify indices $i$ with their binary representation $i_1 \ldots i_n \in  \01^n$. Formally, a quantum query corresponds to the following unitary map on $n+1$~qubits:
$$
O_x:\ket{i,b}\rightarrow \ket{i,b\oplus x_i},
$$
where $i\in \{0,\ldots,N-1\}$ and $b\in \01$. Given access to an oracle of the above type, we can make a
phase query of the form $O_{x,\pm} : \ket{i}\rightarrow (-1)^{x_i} \ket{i}$ as follows: start with $\ket{i,1}$ and apply the Hadamard transform to the last qubit to obtain $\ket{i}\ket{-}$ where $\ket{-}= (\ket{0}-\ket{1})/\sqrt{2}$. Apply $O_x$ to $\ket{i}\ket{-}$ to obtain $(-1)^{x_i}\ket{i}\ket{-}$. Finally, apply Hadamard transform to the last qubit to send it back to $\ket{1}$.

We briefly highlight a few quantum query algorithms that we will invoke later.  

\subsubsection{Grover's algorithm} 
Consider the following \emph{(unordered) search problem}.  A database of size~$N$ is modeled as a binary string~$x\in\01^N$. A \emph{solution} in the database is an index~$i$ such that $x_i=1$. The goal of the search problem is to find a solution given query access to~$x$. It is not hard to see that every classical algorithm that solves the search problem needs to make $\Omega(N)$ queries in the worst case. Grover~\cite{grover:search,bhmt:countingj} came up with a quantum algorithm that finds a solution with high probability using $O(\sqrt{N})$ queries (this is also known to be optimal~\cite{bbbv:str&weak}).

For $N=2^n$, let $D_n=2\ketbra{0^n}{0^n}-\Id$ be the unitary that puts `-1' in front of all basis states except $\ket{0^n}$. The \emph{Grover iterate} $G=H^{\otimes n}D_nH^{\otimes n} O_{x,\pm}$ is a unitary that makes one quantum query. We now describe Grover's algorithm (assuming the number of solutions $|x|=1$).
\begin{enumerate}
\item Start with $\ket{0^n}$.
\item Apply Hadamard transforms to all $n$ qubits, obtaining $\frac{1}{\sqrt{N}} \sum_{i\in \01^n} \ket{i}$.
\item Apply the Grover iterate $G$ $\big \lceil\frac{\pi}{4}\sqrt{N}\big \rceil$ times.
\item Measure the final state to obtain an index $i\in[N]$.
\end{enumerate}
One can show that with high probability the measurement outcome is a solution.  If the number of solutions $|x|\geq 1$ is unknown, then a variant of this algorithm from~\cite{bhmt:countingj} can be used to find a solution with high probability using $O(\sqrt{N/|x|})$ queries.  We will later invoke the following more recent application of Grover's~algorithm. 

\begin{theorem}[{\cite{kothari:oracleidentification},\cite[Theorem~5.6]{linandlin:bombuery}}]
\label{thm:variantofgrover}
Suppose $x\in \01^N$. There exists a quantum algorithm that satisfies the following properties:
\begin{itemize}
\item if $x\neq 0^N$, then let $d$ be the first (i.e., smallest) index satisfying $x_d=1$; the algorithm uses an expected number of $O(\sqrt{d})$ queries to~$x$ and outputs~$d$ with probability at least $2/3$; 
\item if $x=0^N$ then the algorithm always outputs ``no solution'' after $O(\sqrt{N})$ queries.
\end{itemize}
\end{theorem}

\subsubsection{Fourier sampling}\label{sec:fouriersampling}

A very simple but powerful quantum algorithm is \emph{Fourier sampling}. In order to explain it, let us first introduce the basics of Fourier analysis of functions on the Boolean cube (see~\cite{wolf:fouriersurvey,odonnell:analysis} for more).
Consider a function $f:\01^n\to\R$. Its \emph{Fourier coefficients} are~$\widehat{f}(S)=\Exp_x[f(x)\chi_S(x)]$,
where $S\in\01^n$, the expectation is uniform over all $x\in\01^n$, and $\chi_S(x)=~(-1)^{x\cdot S}$ is the \emph{character function} corresponding to~$S$. The Fourier decomposition of $f$ is $ f=\sum_S\widehat{f}(S)\chi_S.$
Parseval's identity says that $\sum_S\widehat{f}(S)^2=\Exp_x[f(x)^2]$.
Note that if $f$ has range $\{\pm 1\}$ then Parseval implies that the squared Fourier coefficients $\widehat{f}(S)^2$ sum to~1, and hence form a probability distribution.
Fourier sampling means sampling an $S$ with probability $\widehat{f}(S)^2$. Classically this is a hard problem, because the probabilities depend on all $2^n$ values of~$f$. However, the following quantum algorithm due to Bernstein and Vazirani~\cite{bernstein&vazirani:qcomplexity} does it exactly using only $1$ query and $O(n)$ gates.
\begin{enumerate}
\item Start with $\ket{0^n}$.
\item Apply Hadamard transforms to all $n$ qubits, obtaining $\frac{1}{\sqrt{2^n}}\sum_{x\in\01^n} \ket{x}$.
\item Query $O_f$,\footnote{Here we view $f \in \{1,-1\}^{2^n}$ as being specified by its truth-table.} obtaining $\frac{1}{\sqrt{2^n}}\sum_x f(x)\ket{x}$. 
\item Apply Hadamard transforms to all $n$ qubits to obtain 
$$
\frac{1}{\sqrt{2^n}}\sum_x f(x)\Big( \frac{1}{\sqrt{2^n}}\sum_S (-1)^{x\cdot S}\ket{S}\Big)=\sum_S \widehat{f}(S)\ket{S}.
$$
\item Measure the state, obtaining $S$ with probability $\widehat{f}(S)^2$.
\end{enumerate}

\subsection{Pretty Good Measurement}
\label{section:pgm}
Consider an ensemble of $m$ $d$-dimensional pure quantum states, $\E=\{(p_i,\ket{\psi_i})\}_{i\in[m]}$, where~$p_i\geq~ 0$ and $\sum_{i\in [m]} p_i=1$. Suppose we are given an unknown state $\ket{\psi_{j}}$ sampled according to the probabilities $\{p_i\}$ and we are interested in maximizing the \emph{average success probability} to identify the given state (i.e., to find $j$). For a POVM $\M=\{M_i\}_{i\in[m]}$, the average success probability is 
$
P_{\M}(\E) = \sum_{i=1}^m p_i\braketbra{\psi_i}{M_i}{\psi_i}.
$

Let $P^{opt}(\E)=\max_{\M} P_{\M}(\E)$ denote the optimal average success probability   of $\E$, maximized over the set of all $m$-outcome POVMs. The  so-called \emph{Pretty Good Measurement} (PGM) is a specific POVM (depending on $\E$), that does \emph{reasonably} well against $\E$. We omit the details of the PGM and state only the crucial properties that we require. Suppose $P^{pgm}(\E)$ is the average success probability in identifying the states in $\E$ using the PGM, then 
$$
P^{opt}(\E) \geq P^{pgm}(\E)\geq  P^{opt}(\E)^2,
$$
 where the second inequality was shown by Barnum and Knill \cite{barnum:pgmmeasurement}.  For an ensemble $\E=\{(p_i,\ket{\psi_i})\}_{i\in[m]}$, let $\ket{\psi'_i}=\sqrt{p_i}\ket{\psi_i}$ for $i\in [m]$. Let $G$ be the $m\times m$ Gram matrix for $\{\ket{\psi'_i}\}_{i\in [m]}$, i.e., $G(i,j)=\ip{\psi'_i}{\psi'_j}$ for $i,j\in [m]$. Then one can show that $P^{pgm}(\E)=\sum_{i\in [m]} \sqrt{G}(i,i)^2$  
(see, e.g., \cite{montanaro:distinguishability} or \cite[Section~2.6]{arunachalam:optimalpaclearning}).

\section{Learning models}
\label{sec:introlearningmodels}
In this section we will define the three main learning models that we focus on: 
the  \emph{exact} model of learning introduced by Angluin \cite{angluin:exactmembership}, the \emph{PAC} model of learning introduced by Valiant~\cite{valiant:paclearning}, and the \emph{agnostic} model of learning introduced by Haussler~\cite{haussler:agnosticlearning} and Kearns et al.~\cite{kearns:agnosticlearning}. 

Below, a \emph{concept class} $\Cc$ will usually be a set of functions $c:\01^n\to\01$, though we can also allow functions $c:[N]\to\01$, or treat such a $c$ as an $N$-bit string specified by its truth-table.

\subsection{Exact learning}

\paragraph{Classical exact learning.}In the exact learning model, a learner $\A$ for a concept class $\Cc$ is given access to a \emph{membership oracle} $\MQ(c)$ for the \emph{target concept}  $c\in \Cc$ that $\A$ is trying to learn. Given an input $x\in \01^n$, $\MQ(c)$ returns the label $c(x)$. A learning algorithm $\A$ is an \emph{exact learner} ~for~$\Cc$ if:
\begin{quote}
For every $c\in\Cc$, given access to the $\MQ(c)$ oracle:\\ 
with probability at least $2/3$, $\A$ outputs an $h$ such that $h(x)=c(x)$ for all $x\in \01^n$.\footnote{We could also consider a $\delta$-exact learner who succeeds with probability $1-\delta$, but here restrict to $\delta=1/3$ for simplicity. Standard amplification techniques can reduce this $1/3$ to any $\delta>0$ at the expense of an $O(\log(1/\delta))$ factor in the complexity.}
\end{quote}
This model is also sometimes known as ``oracle identification'': the idea is that $\Cc$ is a set of possible oracles, and we want to efficiently identify which $c\in\Cc$ is our actual oracle, using membership queries to~$c$.

The \emph{query complexity} of $\A$ is the maximum number of invocations of the $\MQ(c)$ oracle which the learner makes, over all concepts $c\in \Cc$ and over the internal randomness of the learner. The \emph{query complexity of exactly learning}~$\Cc$ is the minimum query complexity over all exact learners~for~$\Cc$.\footnote{This terminology of ``learning $\Cc$'' or ``$\Cc$ is learnable'' is fairly settled though slightly unfortunate: what is actually being learned is of course a target concept $c\in\Cc$, not the class $\Cc$ itself, which the learner already knows from the~start.}

Each concept $c:\01^n\rightarrow \01$ can also be specified by its $N$-bit truth-table (with $N=2^n$), hence one may view the concept class $\Cc$ as a subset of $\01^N$. For a given $N$ and $M$, define the \emph{$(N,M)$-query complexity of exact learning} as the maximum query complexity of exactly learning~$\Cc$, maximized over all~$\Cc\subseteq\01^N$ such that $|\Cc|=M$. 

\paragraph{Quantum exact learning.} In the quantum setting, instead of having access to an $\MQ(c)$ oracle, a \emph{quantum exact learner} is given access to a $\QMQ(c)$ oracle, which corresponds to the map $\QMQ(c):\ket{x,b}\rightarrow \ket{x,b\oplus c(x)}$ for $x\in \01^n, b\in \01$. For a given $\Cc,N,M$, one can define the \emph{quantum query complexity of exactly learning}~$\Cc$, and the \emph{$(N,M)$-quantum query complexity of exact learning} as the quantum analogues to the classical complexity measures.

\subsection{Probably Approximately Correct (PAC) learning}
\paragraph{Classical PAC model.} In the PAC model, a learner $\A$ is given access to a \emph{random example oracle} $\PEX(c,D)$, where $c\in \Cc$ is the \emph{target concept} that $\A$ is trying to learn and $D:\01^n\rightarrow [0,1]$ is an unknown probability distribution. When invoked, $\PEX(c,D)$ returns a labeled example $(x,c(x))$ where $x$ is drawn from $D$.
A learning algorithm $\A$ is an \emph{$(\eps,\delta)$-PAC learner} for $\Cc$ if: 
\begin{quote}
For every $c\in\Cc$ and distribution $D$, given access to the $\PEX(c,D)$ oracle:\\ 
with probability at least $1-\delta$, $\A$ outputs an $h$ such that $\Pr_{x\sim D}[h(x)\neq c(x)]\leq \eps$. 
\end{quote}
Note that the learner has the freedom to output an hypothesis $h$ which is not itself in the concept class~$\Cc$. If the learner always produces an $h\in \Cc$, then it is called a \emph{proper} PAC learner.

The \emph{sample complexity} of $\A$ is the maximum number of invocations of the $\PEX(c,D)$ oracle which the learner makes, over all concepts $c\in \Cc$, distributions $D$, and the internal randomness of the learner. The \emph{$(\eps,\delta)$-PAC sample complexity} of a concept class $\Cc$ is the minimum sample complexity over all $(\eps,\delta)$-PAC learners for~$\Cc$.

\paragraph{Quantum PAC model.} The quantum PAC model was introduced by Bshouty and Jackson~\cite{bshouty:quantumpac}. Instead of having access to a $\PEX(c,D)$ oracle, the \emph{quantum PAC learner} has access to a \emph{quantum example oracle} $\QPEX(c,D)$ that produces a \emph{quantum example} 
$$
\sum_{x\in \01^n} \sqrt{D(x)} \ket{x,c(x)}.
$$
Such a quantum example is the natural quantum generalization of a classical random sample.\footnote{We could also allow complex phases for the amplitudes $\sqrt{D(x)}$; however, these will make no difference for the results presented here.}
While it is not always realistic to assume access to such (fragile) quantum states, one can certainly envision learning situations where the data is provided by a coherent quantum process. 

A quantum PAC learner is given access to several copies of the quantum example and performs a POVM, where each outcome is associated with an hypothesis. Its \emph{sample complexity} is the maximum number of invocations of the $\QPEX(c,D)$ oracle which the learner makes, over all distributions~$D$ and over the learner's internal randomness. We define the \emph{$(\varepsilon,\delta)$-quantum PAC sample complexity} of $\Cc$ as the minimum sample complexity over all $(\varepsilon,\delta)$-quantum PAC learners~for~$\Cc$.

Observe that from a quantum example $\sum_x \sqrt{D(x)} \ket{x,c(x)}$, we can obtain $\sum_x \sqrt{D(x)} (-1)^{c(x)} \ket{x}$ with probability~$1/2$: apply the Hadamard transform to the last qubit and measure it. With probability $1/2$ we obtain the outcome~1, in which case the remaining state is $\sum_x \sqrt{D(x)} (-1)^{c(x)} \ket{x}$. If $D$ is the uniform distribution, then the obtained state is exactly the state needed in step~3 of the Fourier sampling algorithm described in Section~\ref{sec:fouriersampling}.

How does the model of quantum examples compare to the model of quantum membership queries? If the distribution~$D$ is known, a membership query can be used to create a quantum example: the learner can create the superposition $\sum_x\sqrt{D(x)}\ket{x,0}$ and apply a membership query to the target concept~$c$ to obtain a quantum example.
On the other hand, as Bshouty and Jackson~\cite{bshouty:quantumpac} already observed, a membership query cannot be simulated using a small number of quantum examples. Consider for example the learning problem corresponding to Grover search, where the concept class $\Cc\subseteq\01^N$ consists of all strings of weight~1. We know that $\Theta(\sqrt{N})$ quantum membership queries are necessary and sufficient to exactly learn the target concept with high probability. However, it is not hard to show that, under the uniform distribution, one needs $\Omega(N)$ quantum examples to exactly learn the target concept with high probability. Hence simulating one membership query requires at least $\Omega(\sqrt{N})$ quantum examples.

\subsection{Agnostic learning}
\paragraph{Classical agnostic model.} In the PAC model one assumes that the labeled examples are generated perfectly according to a target concept $c\in\Cc$, which is often not a realistic assumption.  In the agnostic model, for an unknown distribution $D:\01^{n+1}\rightarrow [0,1]$, the learner is given access to an $\AEX(D)$ oracle. Each invocation of $\AEX(D)$ produces labeled examples $(x,b)$ drawn from the distribution~$D$ (where $x\in \01^n$ and $b\in \01$).  Define the error of $h:\01^n\rightarrow \01$ under $D$ as $\err_D(h)=\Pr_{(x,b)\sim D}[h(x)\neq b]$. When $h$ is restricted to come from a concept class $\Cc$, 
the minimal error achievable is $\opt_D(\Cc)=\min_{c\in \Cc} \{\err_D(c)\}$. A learning algorithm $\A$ is an \emph{$(\eps,\delta)$-agnostic learner} for $\Cc$ if it can produce a hypothesis $h\in\Cc$ whose error is not much worse: 
\begin{quote}
For every distribution $D$ on $\01^{n+1}$, given access to the $\AEX(D)$~oracle:\\ 
with probability at least $1-\delta$, $\A$ outputs an $h\in\Cc$ such that $\err_D(h)\leq \opt_D(\Cc)+\eps$.
\end{quote}
If there exists a $c\in \Cc$ that perfectly classifies every $x$ with label $b$ for all $(x,b)$ such that $D(x,b)>0$, then $\opt_D(\Cc)=0$ and we are in the setting of proper PAC learning. The \emph{sample complexity} of $\A$ is the maximum number of invocations of the $\AEX(D)$ oracle which the learner makes, over all distributions~$D$ and over the learner's internal randomness. The \emph{$(\eps,\delta)$-agnostic sample complexity} of a concept class $\Cc$ is the minimum sample complexity over all $(\eps,\delta)$-agnostic learners for~$\Cc$. 

\paragraph{Quantum agnostic model.} The model of quantum agnostic learning was first studied in~\cite{arunachalam:optimalpaclearning}. 
For a distribution $D:\01^{n+1}\rightarrow [0,1]$, the \emph{quantum agnostic learner} has access to a $\QAEX(D)$ oracle that produces a quantum example $\sum_{(x,b)\in \01^{n+1}} \sqrt{D(x,b)} \ket{x,b}$. A quantum agnostic learner is given access to several copies of the quantum example and performs a POVM at the end. Similar to the classical complexities, one can define \emph{$(\varepsilon,\delta)$-quantum agnostic sample complexity} as the minimum sample complexity over all $(\varepsilon,\delta)$-quantum agnostic learners for $\Cc$.

\section{Results on query complexity and sample complexity}
\label{sec:querycomplexityandsamplecomplexity}
\subsection{Query complexity of exact learning}

In this section, we begin by proving bounds on the quantum query complexity of  exactly learning a concept class $\Cc$ in terms of a combinatorial parameter $\gamma(\Cc)$, which we define shortly, and then sketch the proof of optimal bounds on $(N,M)$-quantum query complexity of exact learning.

Throughout this section, we will specify a concept $c:\01^n\rightarrow \01$ by its $N$-bit truth-table (with $N=2^n$), hence $\Cc\subseteq \01^N$. For a set $S\subseteq~\01^N$, we will use the ``$N$-bit majority string" $\MAJ(S)~\in \01^N$ defined as: $\MAJ(S)_i=1$ iff $|\{s\in S:s_i=1\}|\geq |\{s\in S:s_i=0\}|$.

\begin{definition}(Combinatorial parameter $\gamma(\Cc)$)
Fix a concept class $\Cc\subseteq \01^N$ of size $|\Cc|>1$, and let $\Cc'\subseteq \Cc$. For $i\in [N]$ and $b\in \01$, define 
$$
\gamma'(\Cc',i,b)=\frac{|\{c\in \Cc':c_i=b\}|}{|\Cc'|}
$$ 
as the fraction of concepts in $\Cc'$ that satisfy $c_i=b$. Let $\gamma'(\Cc',i)=\min \{\gamma'(\Cc',i,0),\gamma'(\Cc',i,1)\}$ be the minimum fraction of concepts that can be eliminated by learning $c_i$.  Let 
$$
\gamma'({\Cc'})=\max_{i\in [N]}\{\gamma'(\Cc',i)\}
$$ 
denote the largest fraction of concepts in $\Cc'$ that can be eliminated by a query. Finally, define
$$
\gamma(\Cc)=\min_{\substack{\Cc'\subseteq \Cc,\\|\Cc'|\geq 2}} \gamma'({\Cc'})=\min_{\substack{\Cc'\subseteq \Cc,\\ |\Cc'|\geq 2}} \hspace{3pt}\max_{i\in [N]}  \hspace{3pt} \min_{b\in \01}  \gamma'(\Cc',i,b).
$$
\end{definition} 

\noindent
This complicated-looking definition is motivated by the following learning algorithm. Suppose the learner wants to exactly learn $c\in\Cc$.
Greedily, the learner would query~$c$ on the ``best'' input $i\in [N]$, i.e., the $i$ that eliminates the largest fraction of concepts from $\Cc$ irrespective of the value of $c_i$. Suppose $j$ is the ``best'' input (i.e., $i=j$ maximizes $\gamma'(\Cc,i)$) and the learner queries~$c$ on index~$j$: at least a $\gamma(\Cc)$-fraction of the concepts in $\Cc$ will be inconsistent with the query-outcome, and these can now be eliminated from~$\Cc$. Call the set of remaining concepts $\Cc'$, and note that $|\Cc'|\leq(1-\gamma(\Cc))|\Cc|$.  The outermost $\min$ in $\gamma(\Cc)$ guarantees that there will be another query that the learner can make to eliminate at least a $\gamma(\Cc)$-fraction of the remaining concepts from $\Cc'$, and so on. We stop when there is only one remaining concept left. Since each query will shrink the set of remaining concepts by a factor of at least $1-\gamma(\Cc)$, making $T=O((\log|\Cc|)/\gamma(\Cc))$ queries suffices to shrink $\Cc$ to $\{c\}$.

\subsubsection{Query complexity of exactly learning \boldmath{$\Cc$} in terms of \boldmath{$\gamma(\Cc)$}}
Bshouty et al.~\cite{bshouty:exactlearningoraclesandqueries} showed the following bounds on the classical complexity of exactly learning a concept class $\Cc$ (we already sketched the upper bound above).

\begin{theorem} [{\cite{bshouty:exactlearningoraclesandqueries,servedio&gortler:equivalencequantumclassical}}]
\label{thm:classialexactlearnerupperbound}
Every classical exact learner for concept class $\Cc$ has to use $\Omega(\max\{1/\gamma(\Cc),\log |\Cc|\})$ membership queries. For every $\Cc$, there is a classical exact learner which learns~$\Cc$ using $O(\frac{\log |\Cc|}{\gamma(\Cc)})$ membership queries.
\end{theorem} 

In order to show a polynomial relation between quantum and classical exact learning, Servedio and Gortler~\cite{servedio&gortler:equivalencequantumclassical} showed the following lower bounds.

\begin{theorem} [{\cite{servedio&gortler:equivalencequantumclassical}}]
\label{thm:quantumlowerboundintermsofgamma}
Let $N=2^n$. Every quantum exact learner for concept class $\Cc\subseteq \01^N$ has to make $\Omega(\max\{\frac{1}{\sqrt{\gamma(\Cc)}},\frac{\log |\Cc|}{n}\})$ membership queries.
\end{theorem}

\begin{proofsketch}  We first prove the $\Omega(1/\sqrt{\gamma(\Cc)})$ lower bound. We will use the unweighted adversary bound of Ambainis~\cite{ambainis:lowerboundsj}. One version of this bound says the following:
suppose we have a quantum algorithm with possible inputs ${\cal D}\subseteq\01^N$, and a relation $R\subseteq {\cal D}\times{\cal D}$ (equivalently, a bipartite graph) with the following properties:

\begin{enumerate}
\item Every left-vertex $v$ is related to at least $m$ right-vertices $w$  (i.e., $|\{w\in \De:(v,w)\in R \}|\geq m$).
\item Every right-vertex $w$ is related to at least $m'$ left-vertices $v$ (i.e., $|\{v\in \De:(v,w)\in R \}|\geq m'$).
\item For every $i\in[N]$, every left-vertex $v$ is related to at most $\ell$ right-vertices $w$ satisfying $v_i\neq w_i$.
\item For every $i\in[N]$, every right-vertex $w$ is related to at most $\ell'$ left-vertices $v$ satisfying $v_i\neq w_i$.
\end{enumerate}
Suppose that for every $(v,w)\in R$, the final states of our algorithm on inputs $v$ and $w$ are $\Omega(1)$ apart in trace norm.
Then the quantum algorithm makes $\Omega(\sqrt{mm'/\ell\ell'})$ queries.

Now we want to apply this lower bound to a quantum exact learner for concept class $\Cc$.
We can think of the learning algorithm as making queries to an $N$-bit input string and producing the name of a concept $c\in\Cc$  as output.
Suppose $\Cc'\subseteq\Cc$ is a minimizer in the definition of $\gamma(\Cc)$ (i.e., $\gamma'(\Cc')=\gamma(\Cc)$).
Define $\tilde{c}=\MAJ(\Cc')$.
Note that $\tilde{c}$ need not be in $\Cc'$ or even in $\Cc$, but we can still consider what our learner does on input $\tilde{c}$. 
We consider two~cases: 

{\bf Case~1:} For every $c\in\Cc'$, the probability that the learner outputs $c$ when run on the typical concept $\tilde{c}$, is $<1/2$. In this case we pick our relation $R=\{\tilde{c}\}\times \Cc'$. Calculating the parameters for the adversary bound, we have $m=|\Cc'|$, $m'=1$, $\ell\leq\gamma'(\Cc')|\Cc'|$ (because for every $i$, $\tilde{c}_i\neq c_i$ for a $\gamma'(\Cc',i)$-fraction of the $c\in\Cc'$ and $\gamma'(\Cc',i)\leq \gamma'(\Cc')$ by definition), and $\ell'=1$. Since, for every $c\in\Cc'$, the learner outputs $c$ with high probability on input $c$, the final states on every pair of $R$-related concepts will be $\Omega(1)$ apart. Hence, the number of queries that our learner makes is $\Omega(\sqrt{mm'/\ell\ell'})=\Omega(1/\sqrt{\gamma'(\Cc')})=\Omega(1/\sqrt{\gamma(\Cc)})$ (because $\Cc'$ minimized $\gamma(\Cc)$).

{\bf Case~2:} There exists a specific $c\in\Cc'$ that the learner gives as output with probability $\geq 1/2$ when run on input~$\tilde{c}$.
In this case we pick $R=\{\tilde{c}\}\times (\Cc'\backslash\{c\})$, ensuring that the final states on every pair of $R$-related concepts will be $\Omega(1)$ apart. We now have $m=|\Cc'|-1$, $m'=1$, $\ell\leq\gamma'(\Cc')|\Cc'|$ (for the same reason as in Case~1), and $\ell'=1$.
Since $(|\Cc'|-1)/|\Cc'|=\Omega(1)$, the adversary bound again yields an $\Omega(1/\sqrt{\gamma(\Cc)})$ bound. \medskip

We now prove the $\Omega((\log |\Cc|)/n)$ lower bound by an information-theoretic argument, as follows.  View the target string $c\in\Cc$ as a uniformly distributed random variable. If our algorithm can exactly identify $c$ with high success probability, it has learned $\Omega(\log|\Cc|)$ bits of information about~$c$ (formally, the mutual information between $c$ and the learner's output is $\Omega(\log|\Cc|)$). From Holevo's theorem~\cite{holevo}, since a quantum query acts on only $n+1$ qubits, one quantum query can yield at most~$O(n)$ bits of information about $c$. Hence $\Omega((\log|\Cc|)/n)$ quantum queries are needed.
\end{proofsketch}

Both of the above lower bounds are in fact individually optimal. First, if one takes $\Cc\subseteq\01^N$ to consist of the $N$ functions $c$ for which $c(i)=1$ for exactly one~$i$, then exact learning corresponds to the unordered search problem with 1 solution. Here $\gamma(\Cc)=1/N$, and $\Theta(\sqrt{N})$ queries are necessary and sufficient thanks to Grover's algorithm.
Second, if $\Cc$ is the class of $N=2^n$ linear functions on $\01^n$, $\Cc=\{c(x)=a\cdot x: a\in\01^n\}$, then Fourier sampling gives an $O(1)$-query algorithm (see Section~\ref{sseclinandjuntas}).  In addition to these quantum-classical separations based on Grover and Fourier sampling, in Section~\ref{sseclinandjuntas} we also mention a fourth-power separation between~$Q(\Cc)$ and $D(\Cc)$ due to Belovs~\cite{belovs:learningsymjuntasj}, for the problem of learning certain \emph{$k$-juntas}. 

Combining Theorems~\ref{thm:classialexactlearnerupperbound} and \ref{thm:quantumlowerboundintermsofgamma}, Servedio and Gortler~\cite{servedio&gortler:equivalencequantumclassical} showed that the classical and quantum query complexity of exact learning are essentially polynomially related for every $\Cc$. 

\begin{corollary} [\cite{servedio&gortler:equivalencequantumclassical}]
\label{cor:polynomialrealation}
If concept class $\Cc$ has classical and quantum membership query complexities~$D(\Cc)$ and $Q(\Cc)$, respectively, then $D(\Cc)=O(nQ(\Cc)^3)$. 
\end{corollary}

\subsubsection{\boldmath{$(N,M)$}-query complexity of exact learning}

In this section we focus on the $(N,M)$-quantum query complexity of exact learning. Classically, the following characterization is easy to prove.

\begin{theorem} [{Folklore}]
The $(N,M)$-query complexity of exact learning is $\Theta(\min\{M,N\})$. 
\end{theorem}

In the quantum context, the $(N,M)$-query complexity of exact learning has been completely characterized by Kothari~\cite{kothari:oracleidentification}. Improving on~\cite{ambainis:quantumidentification, ambainis:improvedquantumidentification,ambainisetal:averagecase}, 
he showed the following theorem.

\begin{theorem} [{\cite{kothari:oracleidentification}}]
The $(N,M)$-quantum query complexity of exact learning is $\Theta(\sqrt{M})$ for $M\leq N$ and $\Theta\Big(\sqrt{\frac{N\log M}{\log(N/\log M)+1}} \Big)$  for  $N< M\leq 2^N$.
\end{theorem}

\begin{proofsketch} Consider first the lower bound for the case $M\leq N$. Suppose $\Cc\subseteq \{c\in \01^N:|c|=1\}$ satisfies $|\Cc|=M$. Then, exactly learning $\Cc$ is as hard as the unordered search problem on $M$ bits, which requires $\Omega(\sqrt{M})$ quantum queries. The lower bound for the case $N<M\leq 2^N$ is fairly technical and we refer the reader to \cite{ambainisetal:averagecase}.  

We now sketch the proofs of the upper bound. We use the following notation: for $u\in \01^n$ and $S\subseteq [n]$, let $u_S \in \01^{|S|}$ be the string obtained by restricting $u$ to the indices in~$S$.

We first describe a quantum algorithm that gives a worse upper bound than promised, but is easy to explain. Suppose $\Cc\subseteq \01^N$ satisfies $|\Cc|=M$. Let $c\in \Cc$ be the unknown target concept that the algorithm is trying to learn. The basic idea of the algorithm is as follows: use the algorithm of Theorem~\ref{thm:variantofgrover} to find the first index $p_1\in [N]$ at which $c$ and $\MAJ(\Cc)$ differ. This uses an expected $O(\sqrt{p_1})$ queries to~$c$ (if there is no difference, i.e., $c=\MAJ(\Cc)$, then the algorithm will tell us so after $O(\sqrt{N})$ queries and we can stop). We have now learned the first $p_1$ bits of~$c$. Let $\Cc_1=\{z_{[N]\backslash [p_1]}:z\in \Cc,\hspace{1mm} z_{[p_1-1]}=\MAJ(\Cc)_{[p_1-1]}, \hspace{1mm} z_{p_1}= \overline{\MAJ(\Cc)}_{p_1}\} \subseteq \01^{N-p_1}$ be the set of suffixes of the concepts in $\Cc$ that agree with $\MAJ(\Cc)$ on the first $p_1-1$ indices and disagree with $\MAJ(\Cc)$ on the $p_1$th index. Similarly, let~$c^1=c_{[N]\backslash [p_1]}$ be the ``updated" unknown target concept after restricting $c$ to the coordinates $\{p_1+1,\ldots,N\}$. Next, we use the same idea to find the first index~$p_2\in~[N-p_1]$ such that $(c^1)_{p_2}\neq \MAJ(\Cc_1)_{p_2}$. Repeat this until only one concept is left, and let~$r$ be the number of repetitions (i.e., until $|\Cc_r|=1$).

In order to analyze the query complexity, first note that, for $k\geq 1$, the $k$-th iteration of the procedure gives us $p_k$ bits of $c$. Since the procedure repeated $r$ times, we have $p_1+\cdots+p_r\leq~N$. Second, each repetition in the algorithm reduces the size of $\Cc_i$ by at least a half, i.e., for~$i\geq 2$, $|\Cc_i|\leq |\Cc_{i-1}|/2$. Hence one needs to repeat the procedure at most $r\leq O(\log M)$ times. The last run will use $O(\sqrt{N})$ queries and will tell us that we have learned all the bits of~$c$. It follows that the total number of queries the algorithm makes to $c$ is 
$$ 
\sum_{k=1}^r O(\sqrt{p_k})+O(\sqrt{N})\leq O\left(\sqrt{r}\sqrt{ \sum_{k=1}^r p_k}\right)+O(\sqrt{N})\leq O(\sqrt{N\log M }),
$$ 
where we used the Cauchy-Schwarz inequality and our upper bounds on $r$ and $\sum_i p_i$.\footnote{One has to be careful here because each run of the algorithm of Theorem~\ref{thm:variantofgrover} has a small error probability. Kothari shows how this can be handled \emph{without} the super-constant blow-up in the overall complexity that would follow from naive error reduction.}

This algorithm is an $O(\sqrt{\log(N/\log M)})$-factor away from the promised upper bound. Tweaking the algorithm to save the logarithmic factor uses the following lemma by \cite{generalizedteaching:hegedus}. It shows that there exists an explicit ordering and a string $s^i$ such that replacing $\MAJ(\Cc_i)$ in the basic algorithm leads to faster reduction of $|\Cc_i|$.
 
\begin{lemma} [{\cite[Lemma~3.2]{generalizedteaching:hegedus}}]
\label{lemma:hegedusreduction}
Let  $L\in \N$ and $\Cc\subseteq \01^L$. There exists $s\in \01^L$ and permutation $\pi:[L]\rightarrow [L]$, such that for every $p\in [L]$, we have $|\Cc_p|\leq \frac{|\Cc|}{\max\{2,p\}}$, where $\Cc_p=\{c\in \Cc:c_{\{\pi(1),\ldots,\pi(p-1)\}}=s_{\{\pi(1),\ldots,\pi(p-1)\}}, \hspace{1mm} c_{\pi(p)}\neq s_{\pi(p)}\}$ is the set of strings in $\Cc$ that agree with $s$ at $\pi(1),\ldots,\pi(p-1)$ and disagree at $\pi(p)$.
\end{lemma}

We now describe the final algorithm.
\begin{enumerate}
\item Set $\Cc_1:=\Cc$, $N_1:=N$, and $c^1:=c$.
\item Repeat until $|\Cc_k|=1$
\begin{itemize}
\item  Let $s^k\in \01^{N_k}$ be the string and $\pi^k:[N_k]\rightarrow [N_k]$ be the permutation obtained by applying Lemma~\ref{lemma:hegedusreduction} to $\Cc_k$ (with $L=N_k$)
\item  Search for the first (according to $\pi^k$) disagreement between $s^k$ and $c^k$ using the algorithm of Theorem~\ref{thm:variantofgrover}. Suppose we find a disagreement at index $\pi^k(p_k)\in [N_k]$, i.e., $s^k$ and~$c^k$ agree on the indices $I_k=\{\pi^k(1),\ldots, \pi^k(p_k-1)\}$
\item  Set $N_{k+1}:=N_k-p_k$, $c^{k+1}:=c^{k}_{[N_k]\backslash (I_k\cup \{\pi^k(p_k)\})}$ and \\
$\Cc_{k+1}:=\{ u_{[N_k]\backslash (I_k\cup \{\pi^k(p_k)\})}: u\in \Cc_k,\hspace{1mm} u_{I_k} =s^k_{I_k},\hspace{1mm} {u_{\pi^k(p_k)}}\neq {s^k_{\pi^k(p_k)}}\}$

\end{itemize}
\item Output the unique element of $\Cc_k$.
\end{enumerate}
Let $r$ be the number of times the loop in Step~2 repeats and suppose in the $k$-th iteration we learned $p_k$ bits of~$c$. Then we have $\sum_{k=1}^r p_k\leq N$. The overall query complexity is $T=O(\sum_{k=1}^r \sqrt{p_k})$.
Earlier we had $|\Cc_{k+1}|\leq |\Cc_k|/2$ and hence $r\leq O(\log M)$. But now, from Lemma~\ref{lemma:hegedusreduction} we have $|\Cc_{k+1}|\leq |\Cc_k|/\max \{2,p_k\}$. Since each iteration reduces the size of $\Cc_k$ by a factor of $\max\{2,p_k\}$, we have $\prod_{k=1}^r \max\{2,p_k\}\leq~M$. Solving this optimization problem (i.e.,  $\min T$ s.t.\ $\prod_{k=1}^r \max\{2,p_k\}\leq~M$, $\sum_{k=1}^r p_k\leq N$), Kothari showed
$$
\quad  T=O(\sqrt{M}) \quad \text{ if } M\leq N,  \quad \text{ and }  T=O\Big(\sqrt{\frac{N\log M}{\log(N/\log M)+1}}\Big)  \quad \text{ if } M> N. 
$$

\vspace*{-3em}
\end{proofsketch}

Kothari~\cite{kothari:oracleidentification}, improving upon \cite{servedio&gortler:equivalencequantumclassical,atici&servedio:qlearning}, resolved a conjecture of Hunziker et al.~\cite{hunziker:quantumexactlearning} by showing the following upper bound for quantum query complexity of exact learning.
This is exactly the above algorithm, analyzed in terms of $\gamma(\Cc)$.

\begin{theorem} [\cite{kothari:oracleidentification}] For every concept class $\Cc$, there is a quantum exact learner for $\Cc$ using 
$
O\Big(\sqrt{\frac{1/\gamma(\Cc)}{\log (1/\gamma(\Cc))}} \log |\Cc|\Big)
$
quantum membership queries.
\end{theorem}
Moshkin~\cite{moshkov:conditionaltests} introduced another combinatorial parameter, which Heged{\H u}s~\cite{generalizedteaching:hegedus} called the \emph{extended teaching dimension} \textit{EXT-TD}$(\Cc)$ of a concept class $\Cc$ (we shall not define \textit{EXT-TD}$(\Cc)$ here, see \cite{generalizedteaching:hegedus} for a precise definition). Building upon the work of \cite{moshkov:conditionaltests}, Heged{\H u}s proved the following theorem.

\begin{theorem} [{\cite{moshkov:conditionaltests},\cite[Theorem~3.1]{generalizedteaching:hegedus}}]
\label{thm:classialexactlearnerupperboundintermsofextd}
Every classical exact learner for concept class $\Cc$ has to use $\Omega(\max\{\textit{EXT-TD}(\Cc),\log |\Cc|\})$ membership queries. For every $\Cc$, there is a classical exact learner which learns~$\Cc$ using $O\left(\frac{\textit{EXT-TD}(\Cc)}{\log (\textit{EXT-TD}(\Cc))}\log |\Cc|\right)$ membership queries.
\end{theorem} 

\noindent
Comparing this with Theorem~\ref{thm:classialexactlearnerupperbound}, observe that both $1/\gamma(\Cc)$ and $\textit{EXT-TD}(\Cc)$ give lower bounds on classical query complexity, but the upper bound in terms of $\textit{EXT-TD}(\Cc)$ is better by a logarithmic factor. Also for analyzing quantum complexity, $\textit{EXT-TD}(\Cc)$ may be a superior parameter.

\subsection{Sample complexity of PAC learning}
One of the most fundamental results in learning theory is that the sample complexity of $\Cc$ is tightly determined by a combinatorial parameter called the \emph{VC dimension} of $\Cc$, named after Vapnik and Chervonenkis~\cite{vapnik:vcdimension} and defined as follows.

\begin{definition} (VC dimension)
 Fix a concept class $\Cc$ over $\01^n$. A set $\Se=\{s_1,\ldots,s_t\}\subseteq \01^n$ is said to be \emph{shattered} by a concept class $\Cc$ if  $\{(c({s_1}) \cdots c({s_t})) : c\in \Cc\} =\01^{t}$. In other words, for every labeling $\ell\in \01^{t}$, there exists a $c\in \Cc$ such that $(c({s_1}) \cdots c({s_t}))=\ell$. The VC dimension of $\Cc$ is the size of a largest $\Se\subseteq \01^n$ that is shattered by~$\Cc$.
\end{definition}

 Blumer et al.~\cite{blumer:optimalpacupper} proved that the $(\varepsilon,\delta)$-PAC sample complexity of a concept class $\Cc$ with VC dimension $d$, is lower bounded by $\Omega(d/\eps + \log(1/\delta)/\eps)$,
and they proved an upper bound that was worse by only a $\log(1/\varepsilon)$-factor.
In recent work, Hanneke~\cite{hanneke:optimalpaclower} (improving on Simon~\cite{simon:almostoptimalpac}) got rid of this  logarithmic factor, showing that the lower bound of Blumer et al.\ is in fact optimal. Combining these bounds, we have the following theorem.
\begin{theorem} [\cite{blumer:optimalpacupper,hanneke:optimalpaclower}]
\label{thm:classicalpaclearning}
Let $\Cc$ be a concept class with VC-dim$(\Cc)=d+1$. Then, $\Theta\Big(\frac{d}{\eps}+\frac{\log(1/\delta)}{\eps}\Big) $ examples are necessary and sufficient for an $(\eps,\delta)$-PAC learner for~$\Cc$.
\end{theorem}

This characterizes the number of samples necessary and sufficient for a classical PAC learning in terms of the VC dimension. How many \emph{quantum} examples are needed to learn a concept class $\Cc$ of VC dimension~$d$?
Trivially, \emph{upper} bounds on classical sample complexity imply upper bounds on quantum sample complexity. For some fixed distributions, in particular the uniform one, we will see in the next section that quantum examples can be more powerful than classical examples.

 However, PAC learning requires a learner to be able to learn $c$ under \emph{all possible} distributions~$D$, not just uniform. We showed that quantum examples are \emph{not more powerful} than classical examples in the PAC model, improving over the results of \cite{atici&servedio:qlearning,zhang:improvedvcbound}.

\begin{theorem}[{\cite{arunachalam:optimalpaclearning}}]
\label{thm:optimalpaclowerbound}
Let $\Cc$ be a concept class with VC-dim$(\Cc)=d+1$.  Then, for every $\delta\in (0,1/2)$ and~$\eps\in (0,1/20)$, $\Omega\Big(\frac{d}{\eps} + \frac{1}{\eps}\log \frac{1}{\delta}\Big)$  examples are necessary for an $(\eps,\delta)$-quantum PAC learner for~$\Cc$.
\end{theorem}

\begin{proofsketch}
 The $d$-independent part of the lower bound has an easy proof, which we omit. In order to prove the $\Omega(d/\varepsilon)$ bound, we first define a distribution~$D$ on the shattered set $\Se=\{s_0,\ldots,s_d\} \subseteq \01^n$ as follows: $D(s_0)=1-20\eps$ and $D(s_i)=20\eps/d$ for all~$i\in [d]$. 

A quantum PAC learner is given $T$ copies of the quantum example for an unknown concept~$c$ and needs to output a hypothesis $h$ that is $\eps$-close to $c$. We want to relate this to the state identification problem of Section~\ref{section:pgm}.
In order to render $\eps$-approximation of $c$ equivalent to \emph{identification} of~$c$, we use a $[d,k,r]_2$ linear error-correcting code for $k\geq d/4$, distance $r\geq d/8$,  with generator matrix $M\in \F_2^{d\times k}$ (we know such codes exist if $d$ is a sufficiently large constant). Let $\{Mz:z\in \01^k\}\subseteq\01^d$ be the set of $2^k$ codewords in this linear code; these have Hamming distance $d_H(Mz,My)\geq d/8$ whenever $z\neq y$. For each $z\in \01^k$, consider a concept $c^z$ defined on the shattered set as: $c^z(s_0)=0$ and $c^z(s_i)=(Mz)_i$ for all $i\in [d]$.  Such concepts exist in $\Cc$ because $\Se$ is shattered by~$\Cc$. Additionally, since $r\geq d/8$ we have $\Pr_{s\sim D}[c^z(s)\neq c^y(s)]\geq 5\eps/2$ whenever $z\neq y$. Hence, with probability at least~$1-\delta$, an $(\eps,\delta)$-PAC quantum learner trying to $\eps$-approximate a concept from $\{c^z:z\in\01^k\}$ will exactly \emph{identify} the concept.  

Consider the following state identification problem: for $z\in \01^k$ let $\ket{\psi_z}=\sum_{i\in \{0,\ldots, d\}} \sqrt{D(s_i)} \ket{s_i, c^z(s_i)}$, and $\E=\{(2^{-k},\ket{\psi_z}^{\otimes T})\}_{z\in \01^k}$. Let $G$ be the $2^k \times 2^k$ Gram matrix for this~$\E$. From Section~\ref{section:pgm}, we know that the average success probability of the PGM is $\sum_{z\in \01^k}\sqrt{G}(z,z)^2$.  Before we compute $\sqrt{G}(z,z)$, note that the $(z,y)$-th entry of $G$ is a function~of~$z\oplus y$:
$$
G(z,y)=\frac{1}{2^k}\ip{\psi_z}{\psi_y}^T=\frac{1}{2^k}\Big(1-\frac{20\eps }{d}|M(z\oplus y)|\Big)^T.
$$
The following claim will be helpful in analyzing the $\sqrt{G}(z,z)$ entry of the Gram matrix.

\begin{theorem} [{\cite[Theorem~17]{arunachalam:optimalpaclearning}}]
\label{thm:upperboundonsqrtGram}
For $m\geq 10$, let $f:\01^m\rightarrow \R$ be defined as $f(w)=(1-\beta\frac{|w|}{m})^T$ for some $\beta\in (0,1]$ and $T\in [1, m/(e^3\beta)]$. For $k\leq m$, let $M\in \F_2^{m\times k}$ be a matrix with rank $k$. Suppose matrix $A\in \R^{2^{k}\times 2^{k}}$ is defined as $A(z,y)=(f\circ M)(z\oplus y)$ for $z,y\in \01^k$, then 
$$
 \sqrt{A}(z,z)\leq e^{O(T^2\beta^2/m+\sqrt{Tm\beta})}    \qquad \text{ for all } z\in \01^k.
$$
\end{theorem}
We will not prove this, but mention that the proof of the theorem crucially uses the fact that the $(z,y)$-entry of matrix $A$ is a function of $z\oplus y$, which allows us to diagonalize $A$ easily. Using the theorem and the definition of $P^{pgm}(\E)$ from Section~\ref{section:pgm}, we have
$$
P^{pgm} (\E)=\sum_{z\in \01^k} \sqrt{G}(z,z)^2\stackrel{\text{Thm.}\ref{thm:upperboundonsqrtGram}}{\leq} e^{O(T^2\eps^2/d+\sqrt{Td\eps}-d-T\varepsilon)}.
$$
The existence of an $(\varepsilon,\delta)$-learner implies $P^{opt}(\E)\geq 1-\delta$. Since $P^{opt}(\E)^2\leq P^{pgm}(\E)$, the above quantity is $\Omega(1)$, which implies $T\geq \Omega(d/\eps)$.
\end{proofsketch}

\subsection{Sample complexity of agnostic learning}
The following theorem characterizes the classical sample complexity of agnostic learning in terms of the VC dimension.

\begin{theorem} [\cite{vapnik:agnosticlowerbound,simon:agnosticlowerbound, talagrand:agnosticupperbound}]
\label{thm:classicalagnosticlearning}
Let $\Cc$ be a concept class with VC-dim$(\Cc)=d$.
Then, $\Theta\Big(\frac{d}{\eps^2} + \frac{\log(1/\delta)}{\eps^2}\Big) $ examples are necessary and sufficient for an $(\eps,\delta)$-agnostic learner for $\Cc$.
\end{theorem}

The lower bound was proven by Vapnik and Chervonenkis~\cite{vapnik:agnosticlowerbound} (see also Simon~\cite{simon:agnosticlowerbound}), and the upper bound was proven by Talagrand~\cite{talagrand:agnosticupperbound}. Shalev-Shwartz and Ben-David~\cite[Section~6.4]{shwartz&david:learningbook} 
call Theorems~\ref{thm:classicalpaclearning} and \ref{thm:classicalagnosticlearning} the ``Fundamental Theorem of PAC learning''.

It turns out that the quantum sample complexity of agnostic learning is equal (up to constant factors) to the classical sample complexity. The proof of the lower bound is similar to the proof of the PAC case.

\begin{theorem} [{\cite{arunachalam:optimalpaclearning}}]
Let $\Cc$ be a concept class with VC-dim$(\Cc)=d$. Then, for every $\delta\in (0,1/2)$ and $\eps \in(0,1/10)$,  
$\Omega\Big(\frac{d}{\eps^2} + \frac{1}{\eps^2}\log \frac{1}{\delta}\Big)$ examples are necessary for an $(\eps,\delta)$-quantum agnostic learner for~$\Cc$.
\end{theorem}

\begin{proofsketch} 
We omit the easy proof of the $d$-independent part in the lower bound. In order to prove the $\Omega(d/\varepsilon^2)$ part, similar to the proof of Theorem~\ref{thm:optimalpaclowerbound}, consider a $[d,k,r]_2$ linear code (for $k\geq d/4$, $r\geq d/8$) with generator matrix $M\in \F_2^{d\times k}$. Let $\{Mz:z\in \01^k\}$ be the set of $2^k$ codewords, these have Hamming distance $d_H(Mz,My)\geq d/8$ whenever $z\neq y$. To each $z\in\01^k$ we associate a distribution $D_z$:
$$
D_z(s_i,b)=\frac{1}{d} \Big (\frac{1}{2}+10(-1)^{(Mz)_i+b}\varepsilon \Big), \qquad  \text{for }  (i,b)\in [d]\times \01,
$$ 
where $\Se=\{s_1,\ldots,s_d\}$ is shattered by $\Cc$. Let $c^{z}\in\Cc$ be a concept that labels $\Se$ according to $Mz\in \01^d$. It is easy to see that $c^{z}$ is the minimal-error concept in $\Cc$ w.r.t.\ the distribution $D_z$. Also, any learner that labels $\Se$ according to $\ell\in\01^d$ has an additional error $d_H(Mz,\ell)\cdot 20\varepsilon/d$ compared to $c^z$. Hence, with probability at least $1-\delta$, an $(\eps,\delta)$-quantum agnostic learner will find a labeling~$\ell$ such that $d_H(Mz,\ell)\leq d/20$. Like in the proof of Theorem~\ref{thm:optimalpaclowerbound}, because $Mz$ is a codeword, finding an $\ell$ satisfying $d_H(Mz,\ell)\leq d/20$ is equivalent to \emph{identifying} $Mz$ (and hence~$z$). 

Now consider the following state identification problem: let $\ket{\psi_z}=\sum_{(i,b)\in [d]\times \01} \sqrt{D_z(s_i,b)} \ket{s_i, b}$ for $z\in~\01^k$ and $\E=\{(2^{-k},\ket{\psi_z}^{\otimes T})\}_{z\in \01^k}$. Let $G$ be the Gram matrix for this $\E$. We have
$$
G(z,y)=\ip{\psi_{z}}{\psi_{y}}^T=\frac{1}{2^k}\Big(1-\frac{1-\sqrt{1-100\eps^2}}{d}|M(z\oplus y)|\Big)^T.
$$
Hence, the $(z,y)$-entry of $G$ depends only on $z\oplus y$ and we are in a position to use Theorem~\ref{thm:upperboundonsqrtGram}. Similar to the proof of Theorem~\ref{thm:optimalpaclowerbound}, we obtain
$$
P^{pgm}(\E)=\sum_{z\in \01^k} \sqrt{G}(z,z)^2\stackrel{\text{Thm.}\ref{thm:upperboundonsqrtGram}}{\leq} e^{O(T^2\eps^4/d+\sqrt{Td \eps^2}-d-T\varepsilon^2)}.
$$
This then implies $T=\Omega(d/\eps^2)$ and proves the~theorem.
\end{proofsketch}

We just saw that in sample complexity for the PAC and agnostic models, quantum examples do not provide an advantage. Gavinsky~\cite{gavinsky:predictivelearning} introduced a model of learning called ``Predictive Quantum'' (PQ), a variation of the quantum PAC model. He exhibited a \emph{relational} concept class that is polynomial-time learnable in PQ, while any ``reasonable'' classical model requires an exponential number of labeled examples to learn the class. 

\subsection{The learnability of quantum states}

In addition to learning \emph{classical} objects such as Boolean functions, one may also consider the learnability of \emph{quantum} objects. Aaronson~\cite{aaronson:qlearnability} studied how well a quantum state $\rho$ can be learned from measurement results. We are assuming here that each measurement is applied to $\rho$ itself, so we require as many fresh copies of $\rho$ as the number of measurements used.
The goal is to end up with a \emph{classical description} of a quantum state $\sigma$ that is in some sense \emph{close} to~$\rho$---and which sense of ``closeness'' we require makes a huge difference. Learning such a good approximation of $\rho$  in trace distance is called \emph{state tomography}. 

In general, an $n$-qubit state $\rho$ is a Hermitian $2^n\times 2^n$ matrix of trace~1, and hence described by roughly $2^{2n}$ real parameters. 
For simplicity, let us restrict attention to allowing only two-outcome measurements on the state (Aaronson discusses also the more general case).
Such a measurement is specified by two positive semi-definite operators $E$ and $\Id-E$, and the probability for the measurement to yield the first outcome is $\Tr(E\rho)$.
Since a two-outcome measurement gives at most one bit of information about $\rho$, $\Omega(2^{2n})$ measurement results are necessary to learn a $\sigma$ that is very close to $\rho$ in trace distance or Frobenius norm. Recently it was shown that such a number of copies is also \emph{sufficient}~\cite{odonnell16,haah16}.   

Because of the exponential scaling in the number of qubits, the number of measurements needed for tomography of an arbitrary state on, say, 100 qubits is already prohibitively large.
However, Aaronson showed an interesting and surprisingly efficient PAC-like result: from $O(n)$ measurement results, with measurements chosen i.i.d.\ according to an unknown distribution~$D$ on the set of all possible two-outcome measurements, we can construct an $n$-qubit quantum state~$\sigma$ that has roughly the same expectation value as $\rho$ for ``most'' two-outcome measurements. In the latter, ``most'' is again measured under the same~$D$ that generated the measurements, just like in the usual PAC setting where the ``approximate correctness'' of the learner's hypothesis is evaluated under the same distribution~$D$ that generated the learner's examples. The output state $\sigma$ can then be used to predict the behavior of $\rho$ on two-outcome measurements, and it will give a good prediction for most measurements. Accordingly, $O(n)$ rather than $\exp(n)$ measurement results suffice for ``pretty good tomography'': to approximately learn an $n$-qubit state that is, maybe not close to $\rho$ in trace distance, but still good enough for most practical purposes.  More precisely, Aaronson's result is the following.

\begin{theorem}[\cite{aaronson:qlearnability}]
For every $\delta,\eps,\gamma>0$ there exists a learner with the following property: 
for every distribution $D$ on the set of two-outcome measurements, given $T=n\cdot\poly(1/\eps,1/\gamma,\log(1/\delta))$ measurement results $(E_1,b_1),\ldots,(E_T,b_T)$ where each $E_i$ is drawn i.i.d.\ from $D$ and $b_i$ is a bit with $\Pr[b_i=1]=\Tr(E_i\rho)$, with probability $\geq 1-\delta$ the learner produces the classical description of a state $\sigma$ such that
$$
\Pr_{E\sim D}\left[ |\Tr(E\sigma) - \Tr(E\rho)|>\gamma \right]\leq\eps.
$$
\end{theorem}

Note that the ``approximately correct'' motivation of the original PAC model is now quantified by two parameters $\eps$ and $\gamma$, rather than only by one parameter $\eps$ as before: the output state~$\sigma$ is deemed approximately correct if the value $\Tr(E\sigma)$ has additive error at most $\gamma$ (compared to the correct value $\Tr(E\rho)$), except with probability~$\eps$ over the choice of~$E$.
We then want the output to be approximately correct except with probability $\delta$, like before.
Note also that the theorem only says anything about the \emph{sample} complexity of the learner (i.e., the number $T$ of measurement results used to construct $\sigma$), not about the time complexity, which may be quite bad in general.  

\medskip

\begin{proofsketch}
The proof invokes general results due to Anthony and Bartlett~\cite{anthony&bartlett:interpolation} and Bartlett and Long~\cite{bartlett&long} about learning classes of probabilistic functions\footnote{A probabilistic function $f$ over a set $\Se$ is a function $f:\Se\rightarrow [0,1]$.} in terms of their \emph{$\gamma$-fat-shattering dimension}. This generalizes VC dimension from Boolean to real-valued functions, as follows.
For some set $\E$, let $\Cc$ be a class of functions $f:\E\to[0,1]$. We say that the set $S=\{E_1,\ldots,E_d\} \subseteq\E$ is \emph{$\gamma$-fat-shattered} by $\Cc$ if there exist $\alpha_1,\ldots,\alpha_d\in[0,1]$ such that for all $Z\subseteq[d]$ there is an $f\in~\Cc$~satisfying:
\begin{enumerate}
\item If $i\in Z$, then $f(E_i)\geq\alpha_i+\gamma$.
\item If $i\not\in Z$, then $f(E_i)\leq\alpha_i-\gamma$.
\end{enumerate}
The \emph{$\gamma$-fat-shattering} dimension of $\Cc$ is the size of a largest $S$ that is shattered by $\Cc$.\footnote{Note that if the functions in $\Cc$ have range $\01$ and $\gamma>0$, then this is just our usual VC dimension.}

For the application to learning quantum states, let $\cal E$ be the set of all $n$-qubit measurement operators. The relevant class of probabilistic functions corresponds to the $n$-qubit density~matrices:
$$
\Cc=\{f:\E\to[0,1] \mid \exists~n\mbox{-qubit }\rho\mbox{ s.t.\ }\forall E\in{\cal E}, f(E)=\Tr(E\rho)\}.
$$
Suppose the set $S=\{E_1,\ldots,E_d\}$ is $\gamma$-fat-shattered by~$\Cc$.
This means that for each string $z\in\01^d$, there exists an $n$-qubit state $\rho_z$ from which the bit $z_i$ can be recovered using measurement~$E_i$, with a $\gamma$-advantage over just outputting~1 with probability~$\alpha_i$. Such encodings $z\mapsto\rho_z$ of classical strings into quantum states are called \emph{quantum random access codes}. Using known bounds on such codes~\cite{ambainis:racj}, Aaronson shows that $d=O(n/\gamma^2)$. This upper bound on the $\gamma$-fat-shattering dimension of $\Cc$ can then be plugged into~\cite{anthony&bartlett:interpolation,bartlett&long} to get the theorem. 
\end{proofsketch}

More recently, in a similar spirit of learning quantum objects, Cheng et al.~\cite{chengetal:learningquantummeasurements} studied how many states are sufficient to learn an unknown \emph{quantum measurement}. Here the answer turns out to be linear in the \emph{dimension} of the space, so exponential in the number of qubits. However, learning an unknown quantum state becomes a \emph{dual problem} to their question and using this connection they can reprove the results of Aaronson~\cite{aaronson:qlearnability} in a different way.

\section{Time complexity}
\label{sec:timecomplexity}

In many ways, the best measure of efficient learning is low \emph{time complexity}. While low sample complexity is a necessary condition for efficient learning, the information-theoretic sufficiency of a small sample is not much help in practice if \emph{finding} a good hypothesis still takes much time.\footnote{As is often the case: for many concept classes, finding a polynomial-sized hypothesis~$h$ that is consistent with a given set of examples is NP-hard.}
In this section we describe a number of results where the best quantum learner has much lower time complexity than the best known classical learner.

\subsection{Time-efficient quantum PAC learning}

When trying to find examples of quantum speed-ups for learning, it makes sense to start with the most famous example of quantum speed-up we have: Shor's algorithm for factoring integers in polynomial time~\cite{shor:factoring}. It is widely assumed that classical computers cannot efficiently factor Blum integers (i.e., integers that are the product of two distinct primes of equal bit-length, each congruent to 3 mod 4). 

Prior to Shor's discovery, Kearns and Valiant~\cite{kearns&valiant:blum} had already constructed a concept class $\Cc$ based on factoring, as an example of a simple and efficiently-representable concept class with small VC dimension that is not efficiently learnable. Roughly speaking, each concept $c\in\Cc$ corresponds to a Blum integer~$N$, and a positively-labeled example for the concept reveals~$N$. A concise description of $c$, however, depends on the factorization of $N$, which is assumed to be hard to compute by classical computers. Servedio and Gortler~\cite{servedio&gortler:equivalencequantumclassical} observed that, thanks to Shor's algorithm, this class \emph{is} efficiently PAC learnable by quantum computers. They similarly observed that the factoring-based concept class devised by Angluin and Kharitonov~\cite{angluin&kharitonov:mem} to show hardness of learning even with membership queries, \emph{is} easy to learn by quantum computers.

\begin{theorem}[\cite{servedio&gortler:equivalencequantumclassical}]
If there is no efficient classical algorithm for factoring Blum integers, then 
\begin{enumerate}
\item there exists a concept class that is efficiently PAC learnable by quantum computers but not by classical computers;
\item there exists a concept class that is efficiently exactly learnable from membership queries by quantum computers but not by classical computers.
\end{enumerate}
\end{theorem}

One can construct classical one-way functions based on the assumption that factoring is hard. These functions can be broken (i.e., efficiently inverted) using quantum computers. However, there are other classical one-way functions that we do not known how to break with a quantum computer. Surprisingly, Servedio and Gortler~\cite{servedio&gortler:equivalencequantumclassical} managed to construct concept classes with quantum-classical separation based on any classical one-way function---irrespective of whether that one-way function can be broken by a quantum computer! The construction builds concepts by combining instances of Simon's problem~\cite{simon:power} with the pseudorandom function family that one can obtain from the one-way function.

\begin{theorem}[\cite{servedio&gortler:equivalencequantumclassical}]
If classical one-way functions exist, then there is a concept class $\Cc$ that is efficiently exactly learnable from membership queries by quantum computers but not by classical~computers.
\end{theorem}

\subsection{Learning DNF from uniform quantum examples}\label{ssecDNF}

As we saw in Section~\ref{sec:introlearningmodels}, Bshouty and Jackson~\cite{bshouty:quantumpac} introduced the model of learning from quantum examples. Their main positive result is to show that Disjunctive Normal Form (DNF) formulas are learnable in polynomial time from quantum examples under the uniform distribution. For learning DNF under the uniform distribution from \emph{classical} examples, the best upper bound is quasi-polynomial time~\cite{verbeurgt:learningdnf}. With the added power of \emph{membership queries}, where the learner can actively ask for the label of any $x$ of his choice, DNF formulas are known to be learnable in polynomial time under uniform~$D$~\cite{jackson:dnf}, but polynomial-time learnability \emph{without} membership queries is a longstanding open~problem.

The classical polynomial-time algorithm for learning DNF using membership queries is Jackson's \emph{harmonic sieve} algorithm~\cite{jackson:dnf}. Roughly speaking it does the following. First, one can show that if the target concept $c:\01^n\to\01$ is an $s$-term DNF (i.e., a disjunction of at most $s$ conjunctions of variables and negated variables) then there exists an $n$-bit parity function that agrees with $c$ on a $1/2+\Omega(1/s)$ fraction of the $2^n$ inputs. Moreover, the Goldreich-Levin algorithm~\cite{goldreich&levin} can be used to efficiently \emph{find} such a parity function with the help of membership queries. This constitutes a ``weak learner'': an algorithm to find a hypothesis that agrees with the target concept with probability at least $1/2+1/\poly(s)$. Second, there are general techniques known as ``boosting''~\cite{freund:boostingbymajority} that can convert a weak learner into a ``strong'' learner, i.e., one that produces a hypothesis that agrees with the target with probability $1-\eps$ rather than probability $1/2+1/\poly(s)$. Typically such boosting algorithms assume access to a weak learner that can produce a weak hypothesis under every possible distribution~$D$, rather than just uniform~$D$. The idea is to start with distribution $D_1=D$, and use the weak learner to learn a weak hypothesis $h_1$ w.r.t.~$D_1$. Then define a new distribution $D_2$ focusing on the inputs where the earlier hypothesis failed; use the weak learner to produce a weak hypothesis $h_2$ w.r.t.\ $D_2$, and so on. After $r=\poly(s)$ such steps the overall hypothesis $h$ is defined as a majority function applied to $(h_1,\ldots,h_r)$.\footnote{Note that this is not \emph{proper} learning: the hypothesis~$h$ need not be an $s$-term DNF itself.}
Note that when learning under fixed uniform~$D$, we can only sample the first distribution~$D_1=D$ directly. Fortunately, if one looks at the subsequent distributions $D_2,D_3,\ldots,D_r$ produced by boosting in this particular case, sampling those distributions $D_i$ can be efficiently ``simulated'' using samples from the uniform distribution. Putting these ideas together yields a classical polynomial-time learner for DNF under the uniform distribution, using membership queries.

The part of the classical harmonic sieve that uses membership queries is the Goldreich-Levin algorithm for finding a parity (i.e., a character function $\chi_S$) that is a weak hypothesis. The key to the \emph{quantum} learner is to observe that one can replace Goldreich-Levin by Fourier sampling from uniform quantum examples (see Section~\ref{sec:fouriersampling}). Let $f=1-2c$, which is just $c$ in $\pm 1$-notation. If~$\chi_S$ has correlation $\Omega(1/s)$ with the target, then $\widehat{f}(S)=\Omega(1/s)$ and Fourier sampling outputs that~$S$ with probability $\Omega(1/s^2)$. Hence $\poly(s)$ runs of Fourier sampling will with high probability give us a weak hypothesis. Because the state at step~3 of the Fourier sampling algorithm can be obtained with probability $1/2$ from a uniform quantum example, we do not require the use of membership queries anymore. Describing this algorithm (and the underlying classical harmonic sieve) in full detail is beyond the scope of this survey, but the above sketch hopefully gives the main ideas of the result of~\cite{bshouty:quantumpac}.

\begin{theorem}[\cite{bshouty:quantumpac}]
The concept class of $s$-term DNF is efficiently PAC learnable under the uniform distribution from quantum examples.
\end{theorem}

\subsection{Learning linear functions and juntas from uniform quantum examples}\label{sseclinandjuntas}

Uniform quantum examples can be used for learning other things as well.
For example, suppose $f(x)=a\cdot x$~mod~2 is a linear function over $\mathbb{F}_2$. Then the Fourier spectrum of $f$, viewed as a $\pm 1$-valued function,
has all its weight on $\chi_a$. Hence by Fourier sampling we can perfectly recover~$a$ with $O(1)$ quantum sample complexity and $O(n)$ time complexity.
In contrast, classical learners need $\Omega(n)$ examples to learn~$f$, for the simple reason that each classical example (and even each membership query, if those are available to the learner too) gives at most one bit of information about the target concept.

A more complicated and interesting example is learning functions that depend (possibly non-linearly) on at most $k$ of the $n$ input bits, with $k\ll n$. Such functions are called \emph{$k$-juntas}, since they are ``governed'' by a small subset of the input bits. We want to learn such $f$ up to error~$\eps$ from uniform (quantum or classical) examples.
A trivial learner would sample $O(2^k\log n)$ classical examples and then go over all ${n\choose k}$ possible sets of up $k$ variables in order to find one that is consistent with the sample. This gives time complexity $O(n^k)$.
The best known upper bound on time complexity~\cite{mos:learningjuntas} is only slightly better:
$O(n^{k\omega/(\omega+1)})$, where $\omega\in[2,2.38]$ is the optimal exponent for matrix multiplication.

Time-efficiently learning $k$-juntas under the uniform distribution for $k=O(\log n)$ is a notorious bottleneck in classical learning theory, since it is a special case of DNF learning: every $k$-junta can be written as an $s$-term DNF with $s<2^k$, by just taking the OR over the 1-inputs of the underlying $k$-bit function. In particular, if we want to efficiently learn $\poly(n)$-term DNF from uniform examples (still an open problem, as mentioned in the previous section) then we should at least be able to efficiently learn $O(\log n)$-juntas (also still open).

Bshouty and Jackson's DNF learner from uniform quantum examples implies that we can learn $k$-juntas using $\poly(2^k,n)$ quantum examples and time (for fixed $\eps,\delta$).
At\i c\i\ and Servedio~\cite{atici&servedio:testing} gave a more precise upper bound.

\begin{theorem}[\cite{atici&servedio:testing}]
There exists a quantum algorithm for learning $k$-juntas under the uniform distribution that uses $O(k\log(k)/\eps)$ uniform quantum examples, $O(2^k)$ uniform classical examples, and $O(nk\log(k)/\eps + 2^k\log(1/\eps))$ time.
\end{theorem}

\begin{proofsketch}
The idea is to first use Fourier sampling from quantum examples to find the $k$ variables (at least the ones with non-negligible influence), and then to use $O(2^k)$ uniform classical examples to learn (almost all of) the truth-table of the function on those variables.

\newcommand{\Inf}{{\rm Inf}}
View the target $k$-junta $f$ as a function with range $\pm 1$. Let the \emph{influence} of variable $x_i$ on~$f$ be
$$
\Inf_i(f)=\sum_{S:S_i=1}\widehat{f}(S)^2=\Exp_x\Big[\Big(\frac{f(x)-f(x\oplus e_i)}{2}\Big)^2\Big]=\Pr_x[f(x)\neq f(x\oplus e_i)],
$$
where $x\oplus e_i$ is $x$ after flipping its $i$th bit.
If $S_i=1$ for an $i$ that is not in the junta, then $\widehat{f}(S)=0$. Hence Fourier sampling returns an $S$ such that $S_i=1$ only for variables in the junta. $\Inf_i(f)$ is exactly the probability that $S_i=1$. Hence for a fixed~$i$, the probability that $i$ does \emph{not} appear in~$T$ Fourier samples is
$$
(1-\Inf_i(f))^T\leq e^{-T\hspace{0.6mm} \Inf_i(f)}.
$$
If we set $T=O(k\log(k)/\eps)$ and let $V$ be the union of the supports of the $T$ Fourier samples, then with high probability $V$ contains all junta variables except those with $\Inf_i(f)\ll \eps/k$ (the latter ones can be ignored since even their joint influence is negligible).

Now use $O(2^k\log(1/\eps))$ uniform classical examples.  With high probability, at least $1-\eps/2$ of all $2^{|V|}$ possible settings of the variables in $V$ will appear, and we use those to formulate our hypothesis~$h$ (say with random values for the few inputs of the truth-table that we didn't see in our sample, and for the ones that appeared twice with inconsistent $f$-values). One can show that, with high probability, $h$ will disagree with $f$ on at most an $\eps$-fraction of $\01^n$.
\end{proofsketch}

In a related result, Belovs~\cite{belovs:learningsymjuntasj} gives a very tight analysis of the number of quantum membership queries (though not the time complexity) needed to exactly learn $k$-juntas whose underlying $k$-bit function is symmetric. For example, if the $k$-bit function is OR or Majority, then $O(k^{1/4})$ quantum membership queries suffice. For the case of Majority, $\Theta(k)$ classical membership queries are required, giving a fourth-power separation between quantum and classical membership query complexity of exact learning (see  Corollary~\ref{cor:polynomialrealation}).

\section{Conclusion}
\label{sec:summaryandoutlook}

Quantum learning theory studies the theoretical aspects of quantum machine learning. We surveyed what is known about this area. Specifically
\begin{itemize}
\item {\bf Query complexity of exact learning.} The number of quantum membership queries needed to exactly learn a target concept can be polynomially smaller than the number of classical membership queries, but not much smaller than that.
\item {\bf Sample complexity.} For the distribution-independent models of PAC and agnostic learning, quantum examples give no significant advantage over classical random examples: for every concept class, the classical and quantum sample complexities are the same up to constant factors.
In contrast, for some fixed distributions (e.g., uniform) quantum examples can be much better than classical examples.
\item {\bf Time complexity.} There exist concept classes that can be learned superpolynomially faster by quantum computers than by classical computers, for instance based on Shor's or Simon's algorithm. This holds both in the model of exact learning with membership queries, and in the model of PAC-learning.  If one allows uniform quantum examples, DNF and juntas can be learned much more efficiently than we know how to do classically.
\end{itemize}
We end with a number of directions for future research.
\begin{itemize}
\item Bshouty and Jackson~\cite{bshouty:quantumpac} showed that DNF (i.e., disjunctions of conjunctions of variables and negations of variables) can be efficiently learned from uniform quantum examples. Is the same true of depth-3 circuits? And what about \emph{constant-depth} circuits with unbounded fan-in AND/OR or even threshold gates, i.e., the concept classes AC$^0$ and TC$^0$---might even these be efficiently learnable from uniform quantum examples or even PAC-learnable? The latter is one of Scott Aaronson's ``Ten Semi-Grand Challenges for Quantum Computing Theory''~\cite{aaronson:10challenges}.
Classically, the best upper bounds on time complexity of learning AC$^0$ are quasi-polynomial under the uniform distribution~\cite{lmn:learnability}, and roughly $\exp(n^{1/3})$ in the PAC model (i.e., under all possible distributions)~\cite{klivans&servedio:dnf}; see~\cite{daniely&shalevshwartz:limitdnf} for a recent~hardness~result. 
\item  At\i c\i\ and Servedio~\cite{atici&servedio:qlearning} asked if for every $\Cc$, the upper bound in Corollary~\ref{cor:polynomialrealation} can be improved to $D(\Cc)\leq O(nQ(\Cc)+Q(\Cc)^2)$?
\item Can we characterize the classical and quantum query complexity of exactly learning a concept class~$\Cc$ in terms of the combinatorial parameter $\gamma(\Cc)$, or in terms of the extended teaching dimension of $\Cc$?
\item Can we find more instances of concept classes where quantum examples are beneficial when learning w.r.t.\ some fixed distribution (uniform or otherwise), or some restricted set of distributions?
\item Can we find examples of quantum speed-up in Angluin's~\cite{angluin:exactmembership} model of equivalence queries plus membership queries?
\item Most research in quantum learning theory (and hence this survey) has focused on concept classes of Boolean functions. What about learning classes of \emph{real-valued} or even \emph{vector-valued}~functions?
\item Can we find a \emph{proper} quantum PAC learner with optimal sample complexity, i.e., one whose output hypothesis lies in $\Cc$ itself? Or a \emph{proper} efficient quantum learner for DNF using uniform quantum examples?
\item Can we find practical machine learning problems with a large provable quantum speed-up?
\item Can we use quantum machine learning for ``quantum supremacy'', i.e., for solving some task using 50--100 qubits in a way that is convincingly faster than possible on large classical computers? (See for example \cite{aaronson&chen:quantumsupremacy} for some complexity results concerning quantum~supremacy.)
\end{itemize}

\subsubsection*{Acknowledgments.}
We thank Lane Hemaspaandra for commissioning this survey for the SIGACT News Complexity Theory Column and helpful comments, and Robin Kothari for helpful comments and pointers (including to~\cite{aaronson:10challenges}), and for sending us his 2012 manuscript~\cite{kothari:qclearn}.

\bibliographystyle{alpha}
\bibliography{qcs}


\end{document}